\documentclass[]{aa}
\usepackage{graphicx}

\renewcommand{\phi}{\varphi}
\begin{document}

\title{Extrasolar Trojan Planets close to Habitable Zones}
\titlerunning{Extrasolar Trojan Planets close to Habitable Zones}
\author{R. Dvorak, E. Pilat-Lohinger, R. Schwarz, F. Freistetter}
\authorrunning{ Dvorak et al.}
\offprints{R.\ Dvorak, \email dvorak@astro.univie.ac.at}
\institute{Universit\"atssternwarte Wien, T\"urkenschanzstr. 17, A-1180 Wien, Austria}
\date{Received; accepted}
\abstract{We investigate the stability regions of hypothetical terrestrial
planets around the Lagrangian equilibrium points $L_4$ and $L_5$ in some
specific extrasolar planetary systems. The problem of their stability can be
treated in the framework of the restricted three body problem where the host
star and a massive Jupiter-like planet are the primary bodies and the
terrestrial planet is regarded as being
massless. From these theoretical
investigations one cannot determine the extension of the stable zones around the
equilibrium points. Using numerical experiments we determined their largeness
for three
test systems chosen from the table of the know
extrasolar planets, where a giant planet is moving close to the so-called habitable zone
around the host star in low eccentric orbits. The results show the dependence
of the size and structure of this region, which shrinks significantly with the
eccentricity of the known gas giant.
  
 }

\maketitle{}

\section{Introduction}

When we are dealing with extrasolar planets and especially with
terrestrial planets in habitable zones in such system we are aware that
up to now we only have observational evidence for gas giants (=GG) like Jupiter.
Therefore a dynamical stability study of possible additional planets of the size comparable to the
Earth is a hypothetical one. It will be a task for future space programs to 
proof the existence of such planets. 

There are different possible regions of motion in the so called habitable
zones\footnote{a somewhat rough estimate for it is 'where water could be in liquid form'; for a more precise definition see Kasting et al. (1993), therefore we concentrate
our work on Trojan planets with a small initial eccentricity to minimise the difference of the radiation from the host star} 
for such additional 'terrestrial planets' which depend on the specific dynamical structure of the exoplanetary system. One can distinguish 4 different classes of such orbits:

\begin{enumerate} 

\item C1: when the GG is very close to the star there could exist such stable
  orbits outside the orbit of the GG for time scales long enough to develop a biosphere.

\item C2: when this GG moves far away from the central star (like Jupiter) then stable low eccentric
orbits for additional planets can exist inside an orbit of the GG.

\item C3: when the GG itself moves in the habitable region a terrestrial like 
satellite (like e.g. Titan in the system of Saturn) could be in a stable orbit.

\item C4: when the GG itself moves in the habitable region a Trojan like terrestrial
  planet may move on a stable orbit around one of the Lagrangian equilibrium points $L_4$ or
$L_5$.

\end{enumerate}

Besides the extensive
study of Menou and Tabachnik (2002) concerning the stability of orbits of terrestrial planets in
extrasolar systems there exist investigations for specific systems for the
classes C1 and C2: 
e.g. Asghari et al. (2004),  E\'rdi and P\'al (2003), P\'al and S\'andor (2003), Dvorak et al. (2003a and  2003b). 
In this new work we study the dynamical stability of possible 
terrestrial planets in the 1:1 resonance with the gas giant (class C4) for three selected
known extrasolar
planetary systems\footnote{We do not discuss the problem of habitable regions around a host star
which is still somewhat in contradiction because it depends not only on the
dynamical parameters of the orbits of a planet there, but also on
the astrophysical parameters of the star like the spectral type and the age
(e.g. Lammer et al. (2003))}.
From the cosmogonical point of view one can imagine a possible formation of
two planets in a 1:1 mean motion resonance (e.g. Laughlin and Chambers, 2002)
as result of an interaction with the protoplanetary disc. They also
investigated the stability of two massive planets in such a resonance which
can be on stable orbits up to the masses of Saturn. In another study 
Nauenberg (2002) found an interesting  stable configuration for motions in the
1:1 resonance, namely when the more massive planet has an almost circular
orbit whereas the smaller body has a high eccentric orbit. 
Our interests were focused on
terrestrial planets on low eccentric orbits in the 1:1 resonance, where only
few studies have been undertaken like recently by \'Erdi and S\'andor (2004).

Out from the list of the approximately 120 planets in extrasolar systems
compilated by Jean
Schneider\footnote{ {\it The Extrasolar Planets Encyclopaedia} at http://www.obspm.fr/encycl/encycl.html)}: we
have chosen as test systems HD 114783, HD 23079 and HD 4208 (table 1), where
the presence of a gas giant was found which moves itself  close to the habitable zone:

\begin{table}
\begin{tabular}{llcccc}
\hline
\textbf{Name}&
\textbf{Spectral-}&
\textbf{Mass}&
\textbf{M.sin i}&
\textbf{a}&
\textbf{ecc.}\\
\textbf{}&
\textbf{type}&
\textbf{[$M_{sun}$]}&
\textbf{[$M_{jup}$]}&
\textbf{[AU]}&
\textbf{}\\
\hline
HD 114783&       K0&      0.92&    0.99&    1.20&    0.10\\
HD 23079&      (F8)/G0V&  1.10&    2.61&    1.65&    0.10\\
HD 4208&         G5V&     0.93&    0.80&    1.67&    0.05\\
\hline
\end{tabular}
\caption[]{Characteristics of the three exoplanetary systems with a giant planet
  moving close to the habitable zone.}
\end{table}

\section{Theoretical Considerations}

In the model of the elliptic restricted problem there exist already many investigations
concerning the
stability of the Lagrangian points depending on the mass ratio of the primaries
and the eccentricity of the orbit (e.g. Rabe, 1967). 
Additional work has been done even for cases when
the third mass is not regarded as massless (Marchal, 1991). 
The results of a first order stability analysis in the framework of the
general three-body-problem (loc.cit. p. 46ff) are presented there. With M the
total mass and  $m_1 \leq m_2 \leq m_3$
a mass parameter R was defined as
$R=(m_2+m_3)/M+m_2.m_3/m_1^2+O(m_2^3.m_3/m_1^4)$.
Using these results one can see that in the case of a terrestrial like planet with a relatively
small mass compared to the two primary bodies there is
practically no difference in the stability of the equilibrium points. 
When we take into account the observed eccentricity of the orbit of gas giant,
furthermore the estimated (minimum) mass of the giant planet and a terrestrial planet with the mass of
our Earth it turned out that almost all planetary systems (of the list given by Jean Schneider
(loc.cit.) with a giant planet
have Lagrangian points which are stable. 
But the extension of the stable region around this equilibrium points cannot be
determined with such an analysis. 

For the Jupiter Trojans the
regions of motion have recently been determined via numerical integrations 
and also by
mapping methods by different authors (e.g. S\'andor and \'Erdi (2003), Robutel
(2004)). 
In a simplified circular restricted problem these stability regions can be 
estimated via the Nekoroshev-theorem, where one
finds that most of the known Trojans are in fact inside these stable regions. 
(Eftymiopoulos, 2004). Unfortunately these methods cannot be applied for
extrasolar systems because of the relatively large eccentricity.
A  detailed answer of the extension
can be given only using  the results of numerical simulations of each
extrasolar planetary system under consideration. 

\section{Numerical methods}
We have chosen two complementary numerical methods to answer the question
of the largeness of the stability region; 
both use direct numerical integrations of the equations of motion. 
\begin{itemize}
\item  The LIE-integration method with an adaptive step size (Hanslmeier and
  Dvorak, 1984; Lichtenegger, 1984)
\item A Bulirsch-Stoer integration
\end{itemize}

As a first approach we started the computation in the dynamical model of the 
elliptic restricted problem consisting of the central star, the gas giant
and a hypothetical (massless) terrestrial planet.
The two primaries were always started in their periastron position; for the terrestrial
planet we have taken the following grid of initial conditions: the different semimajor axis 
covered approximately $\pm 5\%$ of the 
fixed semimajor axis of the gas giant with $\Delta a = 0.01$ AU. For the synodic longitude we have
chosen the range  from $20^o< \alpha< 140^o$  with a gridsize of 0.2 degrees.

The method of analysis used was on one side the Fast Lyapunov Indicators
(=FLI, Froeschl\'e et al., 1997) and
on the other side the largest value of the eccentricity of the hypothetical
Trojan planet during the integration time  (Maximum Eccentricity Method = MEM). The
integration time was in some cases up to $10^6$ years. Shorter integration
times may show structures which disappear for longer time intervals of
integration. The FLI is a well known chaos indicator; 
the MEM gives us the information
of the evolution of the orbital parameters within the regarded time interval and reports also
escapes from the region of motion. 
Via the eccentricity of the orbit
the temperature difference on the surface of the hypothetical planet between periastron and apoastron of the
terrestrial planet can be determined.
For a stable climate on the hypothetical terrestrial planet which allows a
stable biosphere to develop we estimate that $e<0.2$ during the integration time is a reasonable value.
To ensure the validity of the results we always use the MEM and the FLI.


\section{Global Results}

To take into account possible errors in the determination of the
orbital eccentricity 
we have undertaken three different runs  for three different
eccentricities of the GG, namely the 'observed' value $\pm 0.05$. 
The initial eccentricity of the fictitious planet as well as the elements were
set to zero. In the initial condition diagram, the maximum eccentricity of the
fictitious Trojan planet during the integration time was
plotted as a function of synodic longitude and initial semimajor axis.
We can see in figure 1 how the extension
of the 'stable region' around the Lagrangian point for $e_{GG}=0.05, 0.10
$ and $0.15$ for the planetary system HD 23079 -- as an example --
 shrinks  rapidly with the primaries' eccentricity and almost 
disappears for $e_{GG}=0.15$. The same behaviour can
be observed for the other two systems which we investigated.

An interesting
feature is that in the region close to the Lagrangian point itself the
fictitious planet achieves the largest eccentricity.
As example we show for  HD 114783 for $e_{GG}=0.10$ the curvature of the region with respect to the
eccentricity (figure 4). This 'hat' like structure appears in all 
regions around the points $L_4$ and $L_5$; the height of the 'hat'
depends on the mass of the GG and the eccentricity of its orbit.
For an exact determination more computations are necessary.

\section{HD 23079}
 
HD 23079 is a central star of 1.1 Solar masses with a gas giant with a mass
of 2.61 Jupiter masses which moves with a semimajor axis a=1.65 AU on an orbit with e=0.1
close to the habitable region. We varied the eccentricity of the gas giant in order to see how
the structure of the stable region diminishes (see figure 1) . For the lowest value of $e$ in
the region of stable motion the eccentricity of the fictitious Trojan planet
stays always smaller than $e = 0.1$, for the actual measured value of
$e = 0.1$  the  eccentricity of the fictitious Trojan planet still fulfils the
requirement mentioned above of always being below $e<0.2$; for a larger
value $e = 0.15$ the respective large eccentricities would -- according
to our hypothesis -- not allow conditions for habitability. 

We also determined the FLIs which show quite well
the same behaviour as it was found with the MEM. There we also confirm the
ring of less stable motion around the Lagrangian point (see figure 2). For the actual measured $e = 0.1$ one
can recognize a bar like structure which is not visible with the MEM method.
On the contrary we can see different rings around the center (see middle graph
of figure 1) by using the MEM. This structure is not present in the FLI plot. 

\begin{figure}
\begin{center}
\includegraphics[width=2.5in,angle=270]{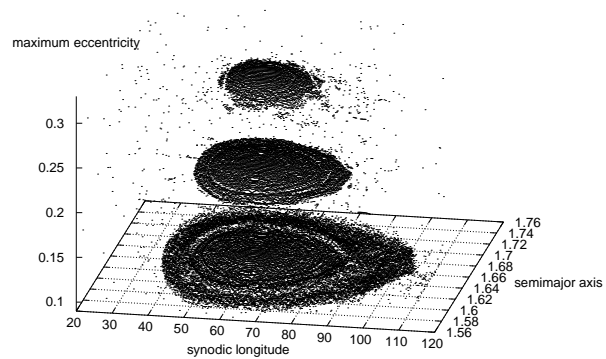}
\caption{Stability region around the system HD 23079 for three different
  values of the eccentricity ($e_{GG}$=0.05, 0.1 and 0.15) of the observed gas giant; for details see in the
  text.}
\end{center}
\end{figure}

\begin{figure}
\begin{center}
\includegraphics[width=2.5in,angle=0]{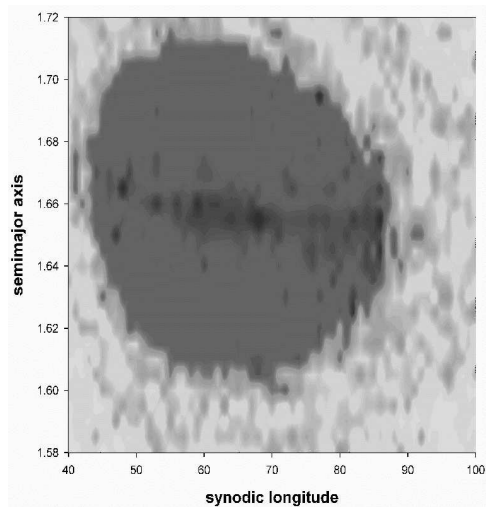}
\caption{FLIs for HD 23079 for $e_{GG}$=0.1, dark regions indicate stable motion.}
\end{center}
\end{figure}

This system was also investigated by another method, namely the RLIs (Relative Lyapunov
Indicators) by \'Erdi and S\'andor (2004) for the actually determined eccentricity
of the primary's orbit. A comparison of the respective results show the good
agreement of all three results.

\section{HD 4208}

HD 4208 is a solar type star with 0.93 Solar masses; the gas giant with a mass
of 0.8 Jupiter masses is orbiting the central star with a=1.67 AU on an almost
circular orbit (e=0.05) which is always within the habitable zone. Again we
varied this eccentricity to see how this changes the size of the
stability region. For the actual value a large area around the Lagrangian
point
is stable and contains orbits for the fictitious Trojan planet which
stays always close of being circular. For $e=0.1$ also orbits for
'habitable' planets would be possible, for $e=0.15$ the large values of the
eccentricity would lead the Trojan planet far out of the habitable zone in the
apoastron and the periastron position. We also checked the region around $L_5$
which turned out to be of the same size as the other equilibrium 
point\footnote{There are about 1054 $L_4$ Jupiter Trojans but only some 628 $L_5$
  Trojans, a fact which is not yet understood}.
Furthermore the comparison with results achieved with the FLIs for the actual
value of $e$ shows a quite good agreement.

\begin{figure}
\begin{center}
\includegraphics[width=2.5in,angle=270]{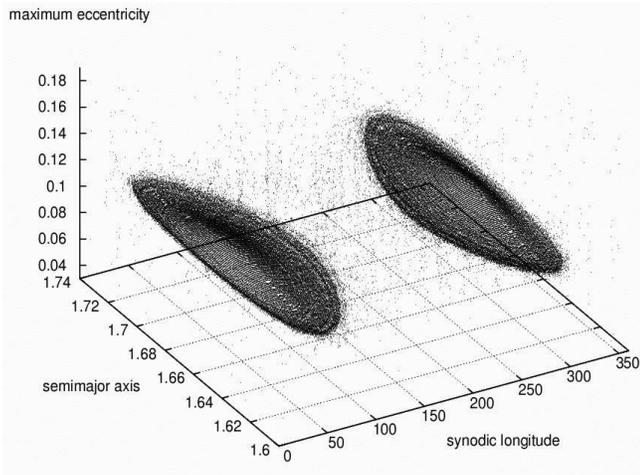}
\caption{Stability region around the system HD 4208; caption like in figure 1.}
\end{center}
\end{figure}

\section{HD 114783}

HD 114783 is a K0 star with almost Solar mass (0.92) which hosts a gas giant of
0.99 Jupiter masses on an eccentric orbit (e=0.1) with a semimajor axis $a=1.2
$ AU. We show the results of the MEM in figure 4, where we can see the hat like structure
with a maximum eccentricity of the orbits around $L_4$ close to the equilibrium
point.
The stable zone itself extends in the synodic longitude $30^o \le \lambda \le
110^o$ and in the semimajor axis from $1.16$ AU $\le a \le  1.24$ AU.

\begin{figure}
\begin{center}
\includegraphics[width=2.5in,angle=270]{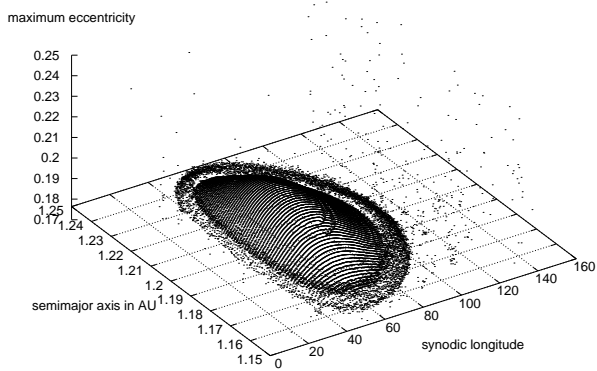}
\caption{Stability region around the system HD 114783 for $e_{GG}=0.1$;
  caption like in figure 1.}
\end{center}
\end{figure}


\section{Conclusions:}

For all three extrasolar planetary systems investigated in this study there is a good
chance for Trojan planets in the 1:1 resonance with the GG to
stay on stable orbits with moderate eccentricities. 
This first stability study will be continued by a analysis
of all possible exosolar systems with a giant planet in the habitable zone
separately.
The aim is then to find how the extension of the stable region around the
equilibrium points depends on one hand on the eccentricity and on the other
hand on the mass of the GG involved.

It may be that future observations of such systems using transits will provide
quite interesting light curves 
(Jean Schneider, 2004) when one recognizes, that a Trojan has a very special
orbit around the equilibrium point which consists of two well distinguished
periods; for the Jupiter Trojans they are about 12 and 160
years.

\begin{acknowledgements}
R. Schwarz and F. Freistetter want to acknowledge the support by the Austrian
FWF (Project P16024) and E. Pilat-Lohinger for the Hertha Firnberg Project T122.
\end{acknowledgements}

\end{document}